\documentclass[traditabstract]{aa}
\usepackage{graphicx}
\usepackage{stfloats}                                                   % to include figure spanning two columns at the bottom of a page
\usepackage{txfonts}
\usepackage{color}
\newcommand{\tablenotea}[1]{\parbox{ 8.8cm}{\indent \footnotesize{#1}}}
\newcommand{\tablenoteb}[1]{\parbox{16.5cm}{\indent \footnotesize{#1}}}
\newcommand{\tablenotec}[1]{\parbox{10.5cm}{\indent \footnotesize{#1}}}
\newcommand{\nature}{Nature}                                     % Nature
\newcommand{\science}{Science}                                 % Science
\newcommand{\jms}{J. Mol. Spec.}                               %
\newcommand{\emp}{Earth Moon Planets}                   %
\newcommand{\mps}{Meteor. Planet. Sci.}                    %
\newcommand{\jsrt}{J. Quant. Spec. Radiat. Transf.} %

\begin{document}
\title{Molecular observations of comets C/2012 S1 (ISON) and C/2013 R1 (Lovejoy): HNC/HCN ratios and upper limits to PH$_3$\thanks{Based on observations carried out with the IRAM 30m Telescope. IRAM is supported by INSU/CNRS (France), MPG (Germany) and IGN (Spain).}}
\titlerunning{Molecular observations of comets ISON and Lovejoy}
\authorrunning{Ag\'undez et al.}

\author{
M. Ag\'undez\inst{1},
N. Biver\inst{2},
P. Santos-Sanz\inst{3},
D. Bockel\'ee-Morvan\inst{2}, and
R. Moreno\inst{2}}

\institute{
Univ. Bordeaux, LAB, UMR 5804,
F-33270, Floirac, France \and
LESIA, Observatoire de Paris, CNRS, UPMC, Universit\' e Paris-Diderot, 5 place Jules Janssen, F-92195 Meudon, France \and
Instituto de Astrof\'isica de Andaluc\'ia - CSIC, Glorieta de la Astronom\'ia s/n, E-18008, Granada, Spain
}

\date{Received; accepted}

% \abstract{}{}{}{}{}
% 5 {} token are mandatory

\abstract
% context heading (optional)
% {} leave it empty if necessary
{We present molecular observations carried out with the IRAM 30m telescope at wavelengths around 1.15 mm towards the Oort cloud comets C/2012 S1 (ISON) and C/2013 R1 (Lovejoy) when they were at $\sim$0.6 and $\sim$1 au, respectively, from the Sun. We detect HCN, HNC, and CH$_3$OH in both comets, together with the ion HCO$^+$ in comet ISON and a few weak unidentified lines in comet Lovejoy, one of which might be assigned to methylamine (CH$_3$NH$_2$). The monitoring of the HCN $J$ = 3-2 line showed a tenfold enhancement in comet ISON on November 14.4 UT due to an outburst of activity whose exact origin is unknown, although it might be related to some break-up of the nucleus. The set of CH$_3$OH lines observed was used to derive the kinetic temperature in the coma, 90 K in comet ISON and 60 K in comet Lovejoy. The HNC/HCN ratios derived, 0.18 in ISON and 0.05 in Lovejoy, are similar to those found in most previous comets and are consistent with an enhancement of HNC as the comet approaches the Sun. Phosphine (PH$_3$) was also searched for unsuccessfully in both comets through its fundamental 1$_0$-0$_0$ transition, and 3$\sigma$ upper limits corresponding to PH$_3$/H$_2$O ratios 4-10 times above the solar P/O elemental ratio were derived.}
% aims heading (mandatory)
{}
% methods heading (mandatory)
{}
% results heading (mandatory)
{}
% conclusions heading (optional), leave it empty if necessary
{}


\maketitle

\section{Introduction}

Radio spectroscopic observations of comets during their visit to the inner solar system have allowed to detect a wide variety of molecules in their coma (e.g., \cite{boc2002} 2002). These observations have provided significant constraints on the chemical nature of comets coming from the two main solar system reservoirs, the Oort cloud and the Kuiper belt, whose composition is expected to reflect to some extent that of the regions of the protosolar nebula where they were once formed.

Two bright comets coming from the Oort cloud approached the Sun in late 2013, allowing us to perform sensitive radio spectroscopic observations and to probe their volatile content. C/2012 S1 (ISON) --hereafter ISON-- was discovered on September 2012 at 6.3 au from the Sun using a 0.4-m telescope of the International Scientific Optical Network (\cite{nev2012} 2012). It is a sungrazing comet, which at perihelion, on 2013 November 28.8 UT, passed at just 0.012 au from the Sun (MPEC 2013-Q27). Its orbital elements are consistent with a dynamically new comet, with fresh ices not previously irradiated by sunlight. A worldwide observational campaign has extensively followed this comet from heliocentric distances beyond 4 au (\cite{oro2013} 2013; \cite{li2013} 2013) to disappearance around perihelion (\cite{kni2014} 2014). C/2013 R1 (Lovejoy) --hereafter Lovejoy-- was discovered in September 2013 at $r_h$ = 1.94 au by Terry Lovejoy using a 0.2-m telescope (\cite{gui2013} 2013). This comet reached perihelion on 2013 December 22.7 UT.  According to its orbital elements (MPEC 2014-D13), this is not its first perihelion passage.

In this Letter we report IRAM 30m spectroscopic observations of the comets ISON and Lovejoy carried out when they were at heliocentric distances of $\sim$0.6 and $\sim$1 au, respectively.

\section{Observations}

The observations of comets ISON and Lovejoy were carried out with the IRAM 30m telescope during the period 13-16 November 2013. At these dates (before perihelion for both comets) ISON spanned a heliocentric distance of 0.67-0.58 au and a geocentric distance of 0.93-0.89 au while Lovejoy was at 1.09-1.06 au from the Sun and 0.43-0.41 au from the Earth. The position of the comets was tracked using the orbital elements from JPL Horizons\footnote{See \texttt{http://ssd.jpl.nasa.gov/horizons.cgi}}.

The EMIR 230 GHz dual polarization receiver (\cite{car2012} 2012) and the FTS spectrometer (\cite{kle2012} 2012) were used to obtain spectra in the frequency ranges 249.0-256.7 GHz and 264.7-272.4 GHz with a spectral resolution of 0.2 MHz ($\sim$0.23 km s$^{-1}$ if expressed as equivalent radial velocity). Important molecular lines such as HCN $J$ = 3-2, HNC $J$ = 3-2, HCO$^+$ $J$ = 3-2, PH$_3$ 1$_0$-0$_0$, and various CH$_3$OH rotational transitions fall within the spectral range covered. Most of the observations were carried out using the wobbler-switching observing mode, with the secondary mirror nutating by $\pm$90$''$ at a rate of 0.5 Hz. Pointing and focus were regularly checked on Mars and nearby quasars. Weather conditions were rather poor during 13 and 14 November, with 6-12 mm of precipitable water vapour (pwv), quite good during 15 November (pwv 1-3 mm), and excellent during 16 November (pwv $<$ 1 mm). The half-power beam width (HPBW) of the IRAM 30m telescope at the observed frequencies ranges from 8.9 to 9.8$''$, and the pointing error is typically lower than 2$''$. Line intensities were converted from antenna temperature $T_A^*$ to main beam brightness temperature $T_{\rm mb}$ by dividing by $B_{\rm eff}$/$F_{\rm eff}$ (e.g., \cite{kra1997} 1997), where $B_{\rm eff}$ is in the range 0.50-0.54 and $F_{\rm eff}$ is 0.88 at the observed frequencies. The data were reduced using the software GILDAS\footnote{See \texttt{http://www.iram.fr/IRAMFR/GILDAS}}.

The bright HCN $J$ = 3-2 line was observed to monitor cometary activity and to locate the position of maximum molecular emission. In both comets the maximum intensity of the HCN $J$ = 3-2 line was found slightly offset from the presumed position of the comet nucleus, at offsets, in (RA, Dec), of ($-8'', +4''$) for ISON (i.e., in the direction of the tail) and of ($0'', +5''$) in the case of Lovejoy. Due to the more favourable weather conditions during 15 and 16 November, the highest quality data for both comets were acquired during these dates, on November 15.4 UT for ISON and November 16.4 UT for Lovejoy. The $T_{\rm mb}$ rms noise levels reached, per 0.2 MHz channel, were 0.013-0.017 K for ISON and 0.008-0.010 K for Lovejoy, after averaging the two polarizations.

\section{HCN monitoring} \label{sec:hcn-monitoring}

\begin{figure}
\centering
\includegraphics[angle=0,width=\columnwidth]{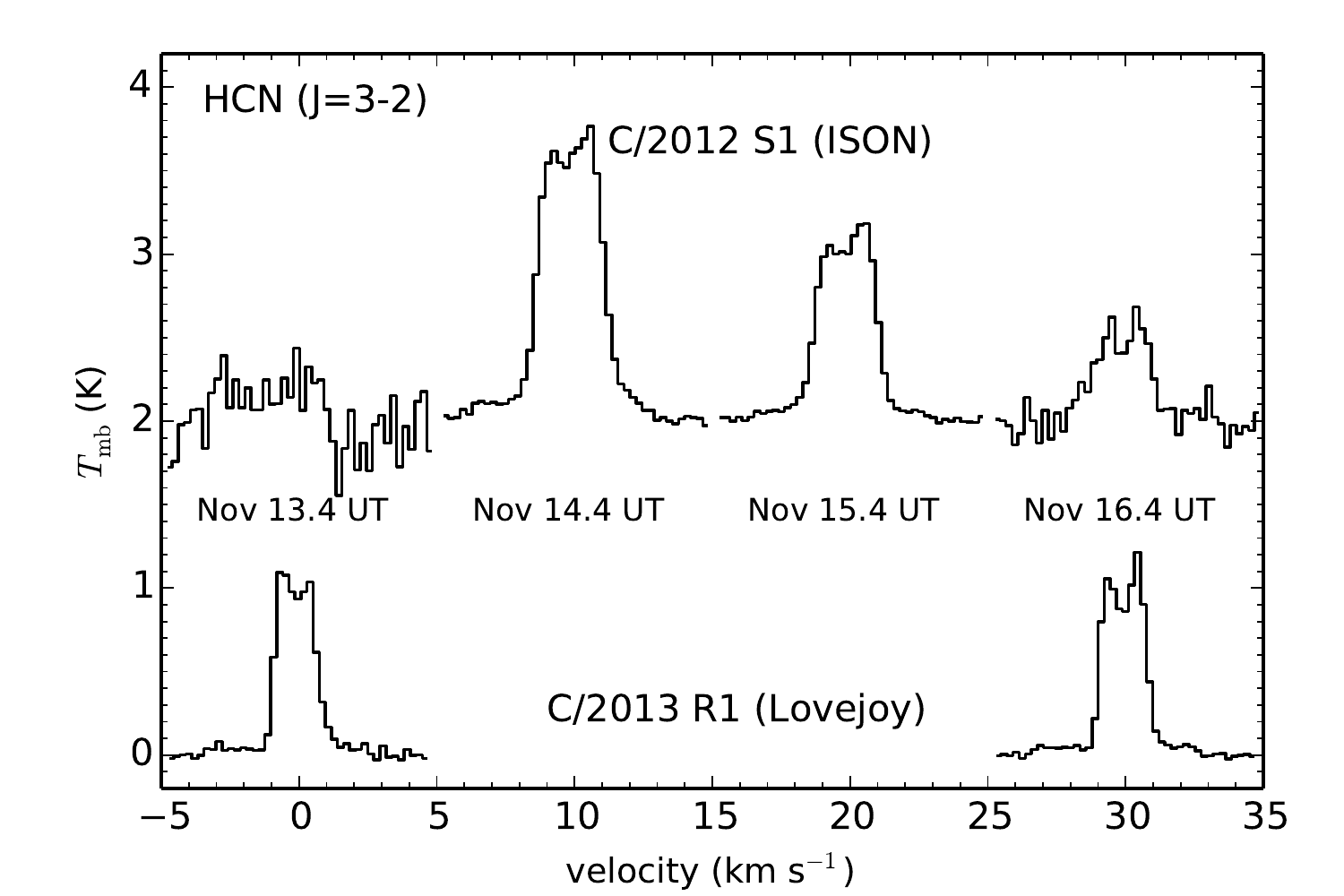}
\caption{Monitoring of the HCN $J$ = 3-2 line on ISON and Lovejoy during November 13.4-16.4 UT. Lines are shifted in velocity by multiple integers of 10 km s$^{-1}$ and in $T_{\rm mb}$ by 2 K for a better visualization. The line intensity shows strong time variations in ISON but remains nearly constant from Nov. 13.4 to 16.4 UT in Lovejoy.} \label{fig:hcn-monitoring}
\end{figure}

A strong variation of the HCN $J$ = 3-2 intensity was observed in comet ISON during November 13.4-16.4 UT, with a tenfold intensity enhancement from Nov. 13.4 to 14.4 UT and a progressive decline afterwards (see Fig.~\ref{fig:hcn-monitoring} and Table~\ref{table:hcn-monitoring}). The outburst of activity of comet ISON on Nov. 14 was reported by various teams observing at different wavelengths. The dramatic increase of the production rate of HCN reported by \cite{biv2013} (2013) and presented here was matched by enhancements in the production rates of other molecules such as OH, CN, and C$_2$ (\cite{cro2013} 2013; \cite{opi2013a} 2013a), an increase in the visual brightness, and the appearance of wings in the coma, which may suggest that the outburst was caused by some splitting of the nucleus (\cite{boe2013} 2013). Another outburst of activity was reported on Nov. 19 by \cite{opi2013b} (2013b). Whether these outbursts were caused by nucleus splitting, delayed sublimation (e.g., \cite{alt2009} 2009), a change in the orientation of the rotation axis, or some other reason is not clear.

In the case of the comet Lovejoy, the intensity of the HCN $J$ = 3-2 line remained nearly constant (within 10 \%) from Nov. 13.4 to 16.4 UT (see Fig.~\ref{fig:hcn-monitoring} and Table~\ref{table:hcn-monitoring}).

\begin{table}
\caption{HCN $J$ = 3-2 line parameters and HCN production rates} \label{table:hcn-monitoring}
\centering
\begin{tabular}{cccccc}
\hline \hline
Date & $r_h$ & $\Delta$ & $\int T_{\rm mb} d$v $^a$ & $\Delta$v $^b$ & $Q$(HCN) \\
(UT) & (au) & (au) & (K km s$^{-1}$)     & (km s$^{-1}$) & (s$^{-1}$) \\
\hline
\multicolumn{6}{c}{C/2012 S1 (ISON)} \\
\hline
Nov. 13.4 & 0.67 & 0.93 & 0.46(14) & 1.5(6) & 6.5 $\times$ 10$^{25}$ \\
Nov. 14.4 & 0.64 & 0.92 & 4.69(3) & 2.37(5) & 6.0 $\times$ 10$^{26}$ \\
Nov. 15.4 & 0.61 & 0.90 & 2.87(2) & 2.22(4) & 3.6 $\times$ 10$^{26}$ \\
Nov. 16.4 & 0.58 & 0.89 & 1.31(7) & 2.21(10) & 1.5 $\times$ 10$^{26}$ \\
\hline
\multicolumn{6}{c}{C/2013 R1 (Lovejoy)} \\
\hline
Nov. 13.4 & 1.09 & 0.43 & 1.92(3) & 1.52(7) & 6.0 $\times$ 10$^{25}$ \\
Nov. 16.4 & 1.06 & 0.41 & 2.06(2) & 1.69(4) & 6.1 $\times$ 10$^{25}$ \\
\hline
\end{tabular}
\tablenotea{\\
Numbers in parentheses are 1$\sigma$ uncertainties in units of the last digits. $^a$ Additional error due to calibration is estimated to be 10-20 \%. $^b$ Line width measured as FWHM.
}
\end{table}

\section{Observed molecules}

\begin{figure}
\centering
\includegraphics[angle=0,width=\columnwidth]{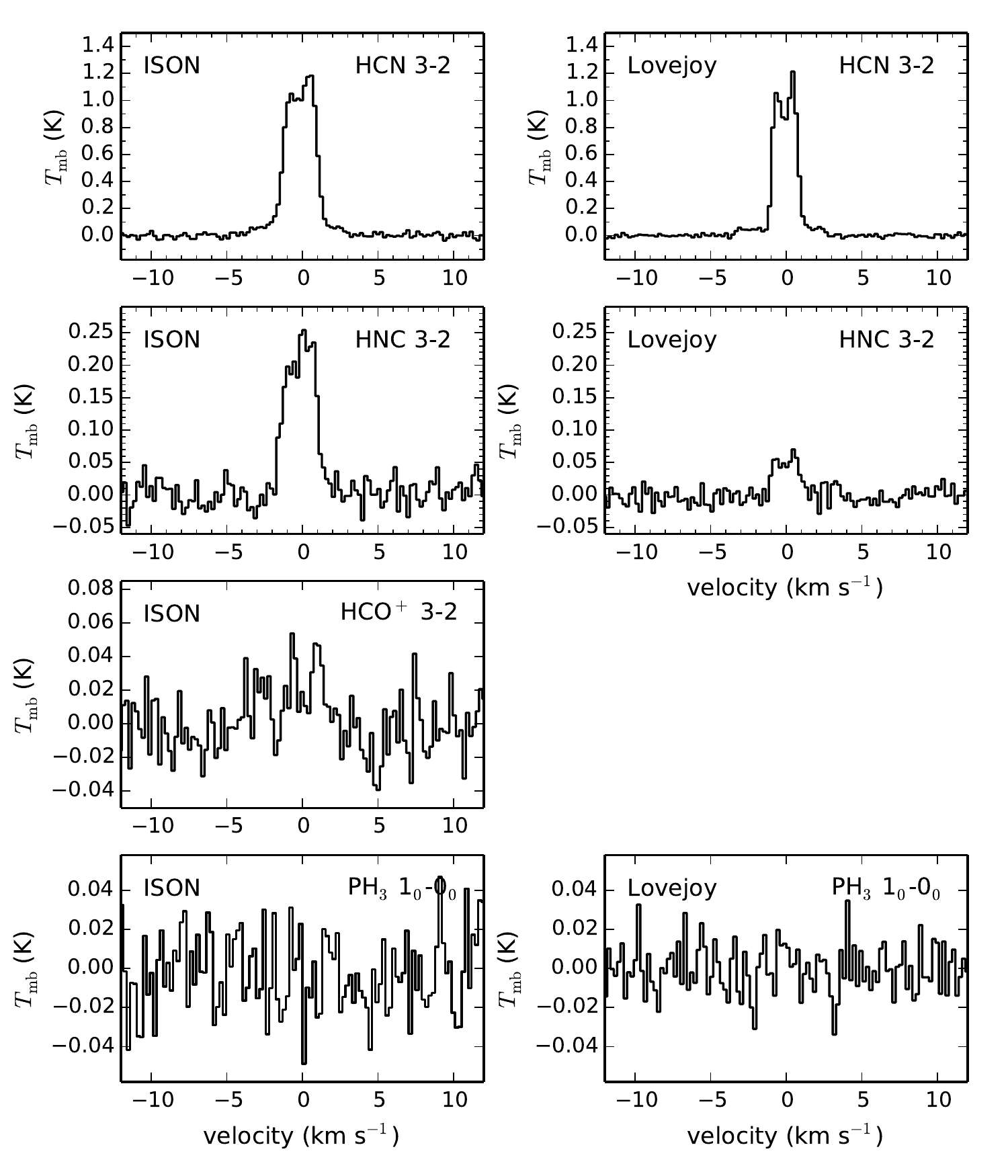}
\caption{Molecular lines observed in comet ISON on Nov. 15.4 UT (left panels) and in comet Lovejoy on Nov. 16.4 UT (right panels). The HCN $J$ = 3-2 and HNC $J$ = 3-2 lines are clearly detected in both comets, the HCO$^+$ $J$ = 3-2 line only in comet ISON, and the PH$_3$ 1$_0$-0$_0$ line is not detected in either of the comets.} \label{fig:lines}
\end{figure}

In addition to the bright $J$ = 3-2 HCN line, some other weaker lines were detected in ISON and Lovejoy within the covered frequency range. We restricted the analysis of lines other than $J$ = 3-2 HCN to the dates of more favourable weather conditions, when acquired spectra were significantly more sensitive, i.e., Nov. 15.4 UT for ISON ($r_h$ = 0.61 au) and Nov. 16.4 UT for Lovejoy ($r_h$ = 1.06 au). A plethora of CH$_3$OH lines and the $J$ = 3-2 line of HNC were detected in both comets, while the $J$ = 3-2 line of HCO$^+$ was only detected in ISON. The 1$_0$-0$_0$ rotational transition of PH$_3$ at 266.9 GHz was searched for in both comets without success, and here we report upper limits for this P-bearing molecule (see section~\ref{subsec:ph3}). Observed lines are shown in Fig.~\ref{fig:lines}, while line parameters and derived production rates are compiled in Table~\ref{table:lines}. Some weak unidentified lines were also detected in Lovejoy (see section~\ref{subsec:ulines}).

\begin{table*}
\caption{Observed lines and inferred molecular production rates in comets ISON at $r_h$ = 0.61 au and Lovejoy at $r_h$ = 1.06 au} \label{table:lines}
\centering
\begin{tabular}{lccccccccc}
\hline \hline
\multicolumn{1}{c}{Molecule} & \multicolumn{1}{c}{Transition} & \multicolumn{1}{c}{$\nu$}   & \multicolumn{1}{c}{$\int T_{\rm mb} d$v $^a$} & \multicolumn{1}{c}{$\Delta$v $^b$} & \multicolumn{1}{c}{$Q$ $^c$} & & \multicolumn{1}{c}{$\int T_{\rm mb} d$v $^a$} & \multicolumn{1}{c}{$\Delta$v $^b$} & \multicolumn{1}{c}{$Q$ $^c$} \\
                                               &                                                & \multicolumn{1}{c}{(MHz)} & \multicolumn{1}{c}{(K km s$^{-1}$)} & \multicolumn{1}{c}{(km s$^{-1}$)} & \multicolumn{1}{c}{(molecule s$^{-1}$)} & & \multicolumn{1}{c}{(K km s$^{-1}$)} & \multicolumn{1}{c}{(km s$^{-1}$)} & \multicolumn{1}{c}{(molecule s$^{-1}$)} \\
\hline
& & & \multicolumn{3}{c}{C/2012 S1 (ISON), Nov. 15.4 UT} & & \multicolumn{3}{c}{C/2013 R1 (Lovejoy), Nov. 16.4 UT} \\
\cline{4-6}
\cline{8-10}
HCN            & $J$ = 3-2   & 265886.4 &  2.87(2) & 2.22(4) & 3.6 $\times$ 10$^{26}$ & & 2.06(2) & 1.69(4) & 6.1 $\times$ 10$^{25}$ \\
HNC            & $J$ = 3-2   & 271981.1 &  0.57(2) & 2.21(8) & 6.4 $\times$ 10$^{25}$ & & 0.11(2) & 1.80(7) & 2.9 $\times$ 10$^{24}$ \\
HCO$^+$    & $J$ = 3-2   & 267557.6 &  0.09(2) & 2.9(10) & -- & & & & \\
CH$_3$OH $^d$ & 11$_0$-10$_1$ A$^+$ & 250507.0 & 0.132(8) & 1.9(2) & 4.3 $\times$ 10$^{27}$ & & 0.101(7) & 1.5(1) & 1.12 $\times$ 10$^{27}$ \\
PH$_3$ $^e$     & 1$_0$-0$_0$ & 266944.5 & $<$0.036 & & $<$8.8 $\times$ 10$^{26}$ & & $<$0.019 & & $<$5.9 $\times$ 10$^{25}$ \\
\hline
\end{tabular}
\tablenoteb{\\
Numbers in parentheses are 1$\sigma$ uncertainties in units of the last digits. $^a$ Additional error due to calibration is estimated to be 10-20 \%. $^b$ Line width measured as FWHM. $^c$ Production rates are computed using the model by \cite{biv2006} (2006). For PH$_3$ we proceeded similarly as to NH$_3$ in \cite{biv2012} (2012). $^d$ The full list of CH$_3$OH lines detected is given in Table~\ref{table:ch3oh}. $^e$ 3$\sigma$ upper limit.
}
\end{table*}

Line widths are $\sim$2.2 km s$^{-1}$ in ISON and $\sim$1.7 km s$^{-1}$ in Lovejoy (see Table~\ref{table:lines}), which imply outflow velocities of 1.1 and 0.85 km s$^{-1}$, respectively, in good agreement with the typical values expected at their heliocentric distances, 0.85 km s$^{-1}$ ($r_h$/au)$^{-1/2}$ according to \cite{bud1994} (1994). The line profiles of HCN and HNC are quite similar, but that of HCO$^+$ in ISON seems to be distinct within the limited signal-to-noise ratio reached (see Fig.~\ref{fig:lines}). Observations of HCO$^+$ in comet C/1995 O1 (Hale-Bopp) indicate that this product species has a complex and variable spatial and kinematical distribution (\cite{lov1998} 1998; \cite{mil2004} 2004). The interpretation of HCO$^+$ data in ISON is beyond the scope of this Letter as it would require to model the chemistry and dynamics of the coma.

\subsection{Methanol: kinetic temperature in the coma}

A large number of methanol lines were observed in both comets within the 249-267 GHz range, most of them belonging to the $J_3$-$J_2$ series of the A$^{\pm}$ and A$^{\mp}$ torsional states (see Table~\ref{table:ch3oh}). The observed line intensities were used in a non-LTE excitation model (\cite{biv2006} 2006) to derive the gas kinetic temperature in the coma, which resulted in 90 K for ISON at $r_h$ = 0.61 au and 60 K for Lovejoy at $r_h$ = 1.06 au. Production rates derived for methanol imply $Q$(CH$_3$OH)/$Q$(HCN) ratios of 12 and 18 for ISON and Lovejoy, respectively, within the typical range of values observed in other comets (e.g., \cite{cro2009} 2009; \cite{boc2011} 2011).

\subsection{HNC/HCN ratio}

\begin{figure}[b]
\centering
\includegraphics[angle=0,width=\columnwidth]{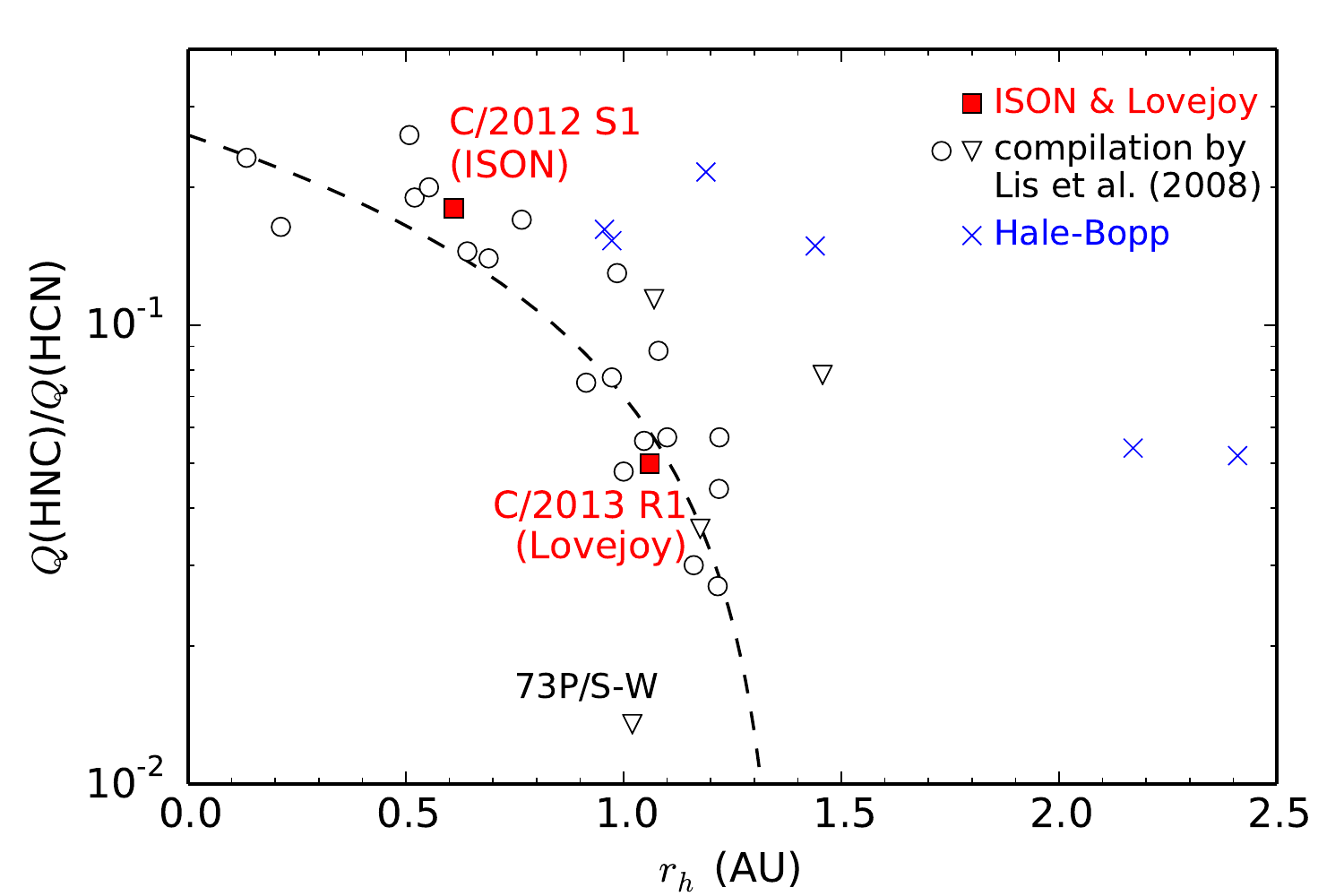}
\caption{HNC/HCN ratios observed in comets as a function of heliocentric distance. Values derived in comets ISON and Lovejoy are indicated as red squares. We also show the values (empty circles) and upper limits (empty triangles) derived in the sample of 14 moderately active comets compiled by \cite{lis2008} (2008). The empirical relation found by these authors is shown as a dashed line. Values derived at various heliocentric distances before perihelion in comet Hale-Bopp (\cite{biv1997} 1997; \cite{irv1998} 1998) are also shown as blue crosses.} \label{fig:hnc2hcn}
\end{figure}

Hydrogen isocyanide (HNC), a metastable isomer of HCN, was first detected in comet C/1996 B2 (Hyakutake) by \cite{irv1996} (1996) and has later on been observed in about a dozen of comets. These observations have served to establish that the HNC/HCN ratio in comets increases as the heliocentric distance reduces, as observed in comet Hale-Bopp (\cite{biv1997} 1997; \cite{irv1998} 1998) and indicated by the empirical correlation found by \cite{lis2008} (2008) in a sample of 14 moderately active comets spanning heliocentric distances in the range 0.1-1.5 au (see Fig.~\ref{fig:hnc2hcn}). It is remarkable that most HNC/HCN ratios derived in comets can be accounted for by a simple expression that solely depends on $r_h$ and not on any intrinsic property of the comet. A few comets show important deviations from this simple relation, however. For example, in the case of 73P/Schwassmann-Wachmann (fragment B) the sensitive upper limit obtained at $r_h$ $\sim$1 au (\cite{lis2008} 2008) points to a HNC/HCN ratio well below that of other comets at similar heliocentric distances. Moreover, for Hale-Bopp the HNC/HCN ratio derived at heliocentric distances beyond 2 au is much higher than expected from extrapolating the behaviour of most other comets. In any case, the variation of the HNC/HCN ratio with $r_h$ indicates that HNC is not directly released from the cometary nucleus but formed in situ in the coma, although its exact origin is still a mystery. It has been argued that HNC may be formed by thermal degradation of organic polymers or dust grains (\cite{rod2001} 2001; \cite{lis2008} 2008), although the high HNC/HCN ratios observed in Hale-Bopp at large heliocentric distances seem to require a different source of HNC. Formation of HNC by radiative isomerization from HCN (\cite{vil2013} 2013) does not seem efficient enough to explain the observed HNC/HCN ratios.

The HNC/HCN ratio in ISON is 0.18 at $r_h$ = 0.61 au, while in Lovejoy this ratio is significantly lower, 0.05 at $r_h$ = 1.06 au (see Fig.~\ref{fig:hnc2hcn}). These values agree with those found in most previous comets and are consistent with a higher HNC/HCN ratio at shorter heliocentric distances. Moreover, the values derived in ISON and Lovejoy match quite well the empirical linear relation found by \cite{lis2008} (2008) between the HNC/HCN ratio and $r_h$, which strengthens the idea that in most comets the HNC/HCN ratio is mainly controlled by the heliocentric distance and not by intrinsic properties of the comet.

\subsection{PH$_3$ upper limits} \label{subsec:ph3}

Phosphine is most likely one of the major carriers of phosphorus in comets, although detecting it remains challenging. The search for it in the very active comet Hale-Bopp while it was at 3.6 au from the Sun yielded an upper limit to $Q$(PH$_3$) of 3.2 $\times$ 10$^{26}$ molecule s$^{-1}$ (\cite{cro2004} 2004), which corresponds to PH$_3$/CO $<$3.2 $\times$ 10$^{-3}$ at 3.6 au (\cite{biv1997} 1997) and translates into PH$_3$/H$_2$O $<$7.4 $\times$ 10$^{-4}$ at 1 au (\cite{boc2000} 2000), that is, somewhat above the solar P/O elemental ratio of 5.2 $\times$ 10$^{-4}$ (\cite{asp2009} 2009).

Our search for PH$_3$ in comets ISON and Lovejoy results in PH$_3$/HCN ratios $<$2.4 and $<$1, respectively (see Table~\ref{table:lines}). Water production rates for these two comets during the relevant dates are not yet accurately known, but based on preliminary estimates (\cite{bon2013} 2013; \cite{com2013} 2013) $Q$(H$_2$O) $\sim$1-3 $\times$ 10$^{29}$ molecule s$^{-1}$ for ISON on Nov. 15.4 UT, which implies an HCN/H$_2$O ratio in the range 0.12-0.36 \%. Adopting a mean HCN/H$_2$O ratio of 0.2 \% for ISON and Lovejoy, the resulting PH$_3$/H$_2$O ratios are $<$5 $\times$ 10$^{-3}$ and $<$2 $\times$ 10$^{-3}$, respectively. These values are 10 and 4 times higher than the solar P/O elemental ratio and thus do not allow us to conclude whether or not PH$_3$ is the main phosphorus species in these comets.

\subsection{Unidentified lines and tentative assignment} \label{subsec:ulines}

\begin{figure}
\centering
\includegraphics[angle=0,width=\columnwidth]{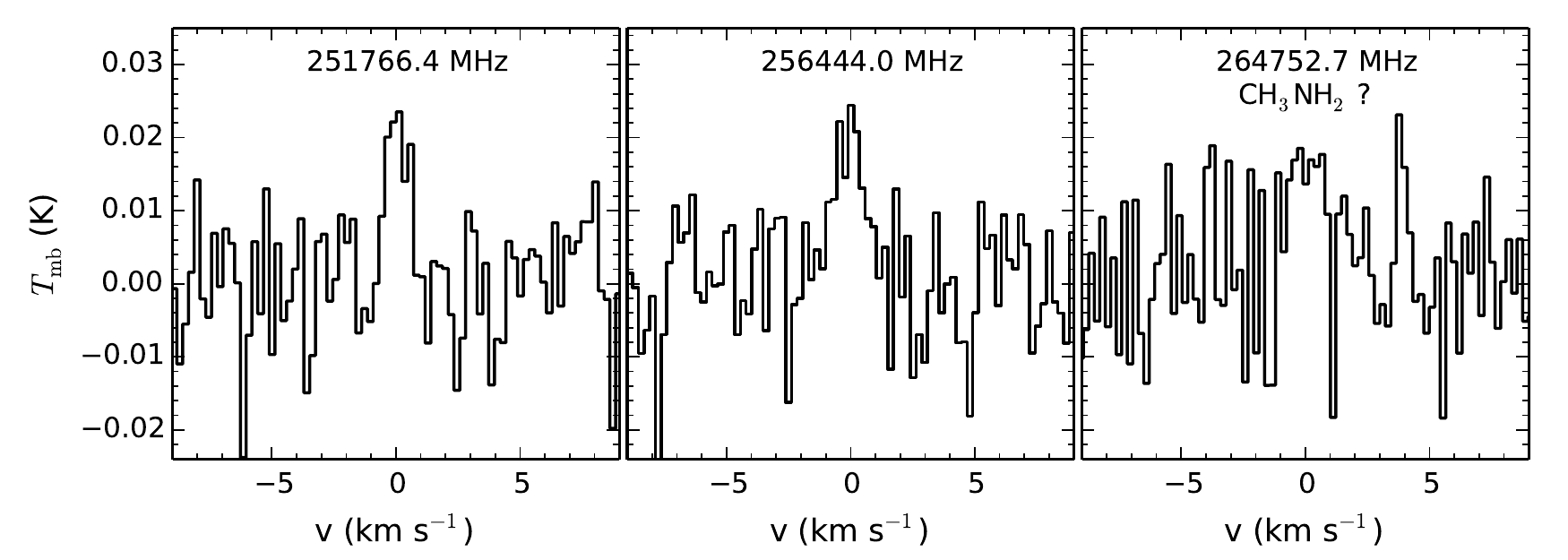}
\caption{Unidentified lines in comet Lovejoy at $r_h$ = 1.06 au.} \label{fig:ulines}
\end{figure}

In addition to the lines of HCN, HNC, and CH$_3$OH observed in comet Lovejoy, a few weak lines with no obvious assignment were also observed in the spectral range covered (see Table~\ref{table:ulines} and Fig.~\ref{fig:ulines}). One of these lines coincides in frequency with the rotational transition 7$_{2,3}$-7$_{1,2}$ of methylamine (CH$_3$NH$_2$), an organic molecule that is a plausible cometary constituent. In fact, this species has been detected in samples of comet Wild 2 returned by the mission Stardust (\cite{gla2008} 2008). Accurate rotational spectroscopic data for this molecule was available just recently (\cite{ily2005} 2005), preventing a search for it in previous comets such as Hale-Bopp (\cite{cro2004} 2004). The inferred production rate of CH$_3$NH$_2$ in Lovejoy is 3-4 $\times$ 10$^{26}$ molecule s$^{-1}$, that is, well above that of HCN. We note, however, that detailed LTE excitation calculations of CH$_3$NH$_2$ in the coma of Lovejoy predict that within the spectral range covered there should be a line at 255997.8 MHz with a similar intensity to the tentatively identified one at 264752.7 MHz. This line is lacking in our observed spectra although it might be within the noise. The low signal-to-noise ratio of the observed line at 264752.7 MHz and the lack of the line at 255997.8 MHz in our data make us to be cautious about the tentative identification of CH$_3$NH$_2$ in Lovejoy.

\section{Summary}

We have carried out IRAM 30m observations of the comets C/2012 S1 (ISON) and C/2013 R1 (Lovejoy) at heliocentric distances of $\sim$0.6 and $\sim$1 au, respectively. We detected HCN, HNC, and CH$_3$OH in both comets, plus the ion HCO$^+$ in ISON and a few weak unidentified lines in Lovejoy, one of which might be assigned to CH$_3$NH$_2$. A tenfold enhancement of the HCN $J$ = 3-2 line was observed in comet ISON within less than 24 h on November 14, indicating an outburst of activity whose origin could be related to nucleus splitting. The large number of CH$_3$OH lines observed was used to derive kinetic temperatures in the coma of 90 and 60 K in ISON and Lovejoy, respectively. The HNC/HCN ratios derived, 0.18 in ISON and 0.05 in Lovejoy, are similar to those found in most previous comets and are consistent with an enhancement of HNC as the comet approaches the Sun. PH$_3$ was also searched for unsuccessfully in both comets so that only upper limits to the PH$_3$/H$_2$O ratio 4-10 times above the solar P/O elemental ratio were derived.

\begin{acknowledgements}

We thank the IRAM 30m staff for their help during the observations and the anonymous referee for a quick and constructive report. P. S.-S. acknowledges financial support by spanish grant AYA2011-30106-C02-01.

\end{acknowledgements}

\onltab{3}{
\begin{table}
\caption{Observed lines of CH$_3$OH in comets ISON at $r_h$ = 0.61 au and Lovejoy at $r_h$ = 1.06 au} \label{table:ch3oh}
\centering
\begin{tabular}{ccccccc}
\hline \hline
\multicolumn{1}{c}{Transition} & \multicolumn{1}{c}{$\nu$}   & \multicolumn{1}{c}{$\int T_{\rm mb} d$v $^a$} & \multicolumn{1}{c}{$\Delta$v $^b$} & & \multicolumn{1}{c}{$\int T_{\rm mb} d$v $^a$} & \multicolumn{1}{c}{$\Delta$v $^b$} \\
                                                & \multicolumn{1}{c}{(MHz)} & \multicolumn{1}{c}{(K km s$^{-1}$)} & \multicolumn{1}{c}{(km s$^{-1}$)}  & & \multicolumn{1}{c}{(K km s$^{-1}$)} & \multicolumn{1}{c}{(km s$^{-1}$)} \\
\hline
& & \multicolumn{2}{c}{C/2012 S1 (ISON)} & & \multicolumn{2}{c}{C/2013 R1 (Lovejoy)} \\
\cline{3-4}
\cline{6-7}
15$_3$-15$_2$ A$^{\mp}$ & 249419.9 & 0.023(6) & 2.4(6) & & & \\
13$_3$-13$_2$ A$^{\mp}$ & 250291.2 & 0.047(8) & 1.4(4) & & & \\
11$_0$-10$_1$ A$^+$       & 250507.0 & 0.132(8) & 1.9(2) & & 0.101(7) & 1.5(1) \\
12$_3$-12$_2$ A$^{\mp}$ & 250635.2 & 0.057(8) & 1.9(2) & & & \\
11$_3$-11$_2$ A$^{\mp}$ & 250924.4 & 0.059(8) & 1.8(3) & & 0.038(7) & 1.6(2) \\
10$_3$-10$_2$ A$^{\mp}$ & 251164.1 & 0.075(10) & 2.4(3) & & 0.034(5) & 1.5(2) \\
9$_3$-9$_2$ A$^{\mp}$    & 251359.9 & 0.070(8) & 1.9(3) & & 0.063(5) & 1.2(1) \\
8$_3$-8$_2$ A$^{\mp}$    & 251517.3 & 0.126(8) & 1.7(2) & & 0.072(5) & 1.4(1) \\
7$_3$-7$_2$ A$^{\mp}$    & 251641.7 & 0.118(8) & 2.2(2) & & 0.088(5) & 1.4(1) \\
6$_3$-6$_2$ A$^{\mp}$    & 251738.5 & 0.139(8) & 2.2(3) & & 0.116(5) & 1.5(1) \\
5$_3$-5$_2$ A$^{\mp}$    & 251811.9 & 0.088(8) & 1.8(2) & & 0.110(5) & 1.5(1) \\
4$_3$-4$_2$ A$^{\mp}$    & 251866.6 & 0.090(10) & 1.8(3) & & 0.098(7) & 1.5(1) \\
5$_3$-5$_2$ A$^{\pm}$    & 251890.9 & 0.129(10) & 2.1(2) & & 0.121(7) & 1.7(1) \\
6$_3$-6$_2$ A$^{\pm}$    & 251895.7 & 0.141(10) & 2.2(2) & & 0.102(7) & 1.4(1) \\
4$_3$-4$_2$ A$^{\pm}$    & 251900.5 & 0.129(10) & 2.1(2) & & 0.100(7) & 1.5(1) \\
3$_3$-3$_2$ A$^{\mp}$    & 251905.8 & 0.084(8) & 1.8(3) & & 0.061(8) & 1.4(3) \\ 
3$_3$-3$_2$ A$^{\pm}$    & 251917.0 &               &            & & 0.062(8) & 1.7(3) \\
7$_3$-7$_2$ A$^{\pm}$    & 251923.6 & 0.108(10) & 2.3(3) & & 0.095(7) & 1.5(1) \\
8$_3$-8$_2$ A$^{\pm}$    & 251984.7 & 0.077(8) & 2.1(3) & & 0.064(7) & 1.7(2) \\
9$_3$-9$_2$ A$^{\pm}$    & 252090.4 & 0.089(10) & 2.1(2) & & 0.051(7) & 1.3(2) \\
10$_3$-10$_2$ A$^{\pm}$ & 252252.8 & 0.097(10) & 2.2(2) & & 0.036(5) & 1.1(1) \\
11$_3$-11$_2$ A$^{\pm}$ & 252485.6 & 0.095(13) & 2.9(6) & & 0.036(7) & 1.2(4) \\
12$_3$-12$_2$ A$^{\pm}$ & 252803.4 & 0.053(12) & 1.8(3) & & 0.016(5) & 0.9(3) \\
2$_{+0}$-1$_{-1}$ E          & 254015.3 &                &           & & 0.061(8) & 2.1(4) \\
6$_{+1}$-5$_{+2}$ E         & 265289.6 &                &           & & 0.060(7) & 1.9(3) \\
5$_{+2}$-4$_{+1}$ E         & 266838.1 & 0.17(2)   & 2.1(2) & & 0.184(9) & 1.5(1) \\
9$_{+0}$-8$_{+1}$ E         & 267403.4 & 0.077(12) & 1.9(3) & & 0.071(7) & 1.6(2) \\
\hline
\end{tabular}
\tablenotec{\\
Numbers in parentheses are 1$\sigma$ uncertainties in units of the last digits. $^a$ Additional error due to calibration is estimated to be 10-20 \%. $^b$ Line width measured as FWHM.
}
\end{table}
}

\onltab{4}{
\begin{table}
\caption{Unidentified lines in comet Lovejoy at $r_h$ = 1.06 au} \label{table:ulines}
\centering
\begin{tabular}{ccccc}
\hline \hline
\multicolumn{1}{c}{$\nu$}  & $\int T_{\rm mb} d$v & $\Delta$v & S/N $^a$ & Note \\
\multicolumn{1}{c}{(MHz)} & (K km s$^{-1}$)         & (km s$^{-1}$) & & \\
\hline
251766.4 & 0.026(6) & 1.0(2) & 4.3 & \\
256444.0 & 0.034(7) & 1.5(3) & 4.9 & \\
264752.7 & 0.029(7) & 1.4(4) & 4.1 & CH$_3$NH$_2$ 7$_{2,3}$-7$_{1,2}$ $?$ \\
\hline
\end{tabular}
\tablenotea{\\
Numbers in parentheses are 1$\sigma$ uncertainties in units of the last digits. $^a$ Line area signal-to-noise ratio.
}
\end{table}
}

\end{document}